\documentclass[conference,compsoc]{IEEEtran}

\ifCLASSOPTIONcompsoc
  \usepackage[nocompress]{cite}
\else
  \usepackage{cite}
\fi

\usepackage{booktabs}
\usepackage{multirow}
\usepackage{amsmath,amssymb,amsfonts}
\usepackage{bm}
\usepackage{graphicx}
\usepackage[caption=false,font=footnotesize]{subfig}
\usepackage{enumitem}
\usepackage{xspace}
\usepackage{microtype}
\usepackage{mathtools}
\usepackage{siunitx}

\begin{document}

\title{Discard the Dross and Select the Essential: Pre-query Sample Selection for Black-box Membership Inference Attacks}


\author{
\IEEEauthorblockN{
Dongdong Zhao\textsuperscript{1},
Jinrong Hu\textsuperscript{1},
Changtian Song\textsuperscript{1},
Jian Chen\textsuperscript{1},
Hongmin Wang\textsuperscript{2},
and Baogang Song\textsuperscript{1,*}
}

\IEEEauthorblockA{
\textsuperscript{1}Wuhan University of Technology\\
Email: zdd@whut.edu.cn, 326026@whut.edu.cn, songchangtian@whut.edu.cn, 321559@whut.edu.cn, 297710@whut.edu.cn
}

\IEEEauthorblockA{
\textsuperscript{2}Bohai University\\
Email: wanghongmin@qymail.bhu.edu.cn
}
}

\maketitle

\begin{abstract}
Black-box membership inference attacks (MIAs) rely on target-model queries to infer whether candidate samples were used for training. However, membership signals are highly non-uniform across samples: some candidate samples support strong member/non-member separability, whereas many others provide little useful signal. Consequently, indiscriminate querying can incur substantial query cost and increase query-induced exposure, with limited marginal benefit for inference. This raises a key question: which candidate samples are worth querying for black-box MIAs?
To address this question, we propose PSS-MIA, a pre-query sample selection framework which can be embedded with any existing MIA methods.  
PSS-MIA proceeds in two stages: it first ranks candidate samples and selects a subset expected to support stronger membership inference, then queries the selected samples and uses the returned outputs for an existing black-box MIA, thereby reducing query cost and query-induced exposure.
In the first stage, we propose Loss-Gap Ranking (LGR), which ranks candidate samples by estimating the strength of their membership signal using loss gaps computed from reference models.
Experiments on CIFAR-10, CIFAR-100, and CINIC-10 with five representative black-box MIA methods demonstrate that PSS-MIA with LGR consistently outperforms all other compared methods. Moreover, under a 0.1\% FPR constraint, PSS-MIA can save at least 83.1\%, 60.6\%, and 80.4\% of the query budget for the three datasets, respectively.

\end{abstract}
\maketitle

\begingroup
\renewcommand{\thefootnote}{\fnsymbol{footnote}}
\makeatletter
\long\def\@makefntext#1{%
  \parindent 1em%
  \noindent
  \hb@xt@1.8em{\hss\@makefnmark}#1%
}
\makeatother
\footnotetext[1]{Corresponding author: Baogang Song.}
\endgroup
\IEEEpeerreviewmaketitle

\begin{figure}[!t]
    \centering
    \includegraphics[width=\columnwidth]{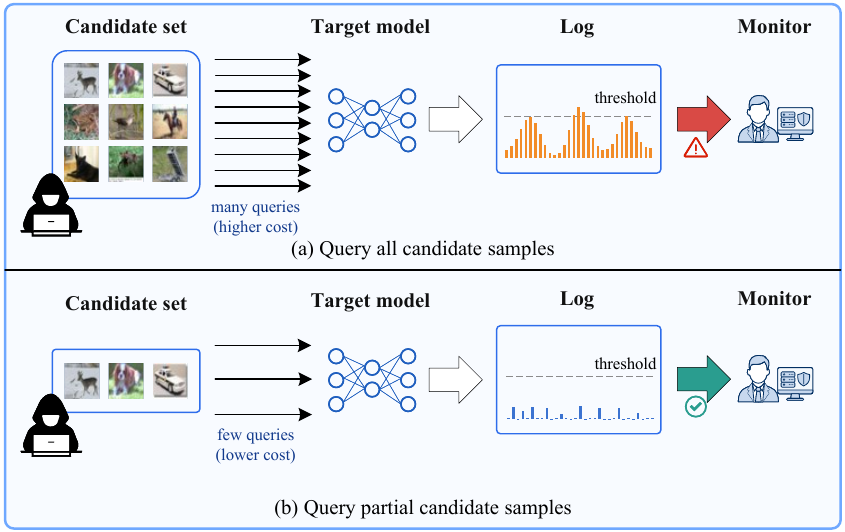}
    \caption{Motivation of pre-query sample selection for black-box MIAs.}
    \label{fig:motivation}
\end{figure}

\section{Introduction}

Membership inference attacks (MIAs) aim to determine whether a given candidate sample was used to train a target model~\cite{shokri2017membership,yeom2018privacy}.
These attacks exploit the fact that trained models often behave differently on training samples and non-training samples, for example, prediction confidence, predicted labels and true-label losses may exhibit different distributions between the two groups~\cite{yeom2018privacy,choquette2021label,carlini2022membership}.
In the black-box setting, the adversary cannot access the training data or model internals and can only infer whether a candidate sample is a member by querying the target model and analyzing the returned outputs~\cite{shokri2017membership,salem2019ml}.
Because this access model matches many deployed prediction services, black-box MIA has become a common tool for assessing membership leakage in machine learning models.

However, existing black-box MIAs typically assume that the adversary first queries candidate samples and then uses the model outputs to infer membership.
Early black-box MIA methods train multiple shadow models to imitate the behavior of the target model and use their prediction vectors to train an attack model~\cite{shokri2017membership,salem2019ml}.
Later methods such as LiRA~\cite{carlini2022membership} use shadow models to estimate IN and OUT distributions and infer membership by comparing the target model output with these distributions~\cite{carlini2022membership,zarifzadeh2024low}.
These methods have substantially advanced black-box membership inference, but they mainly focus on how to infer membership after target model outputs have become available.
This query-all-then-infer practice has clear limitations in practical scenarios.
Target model queries are often costly, especially for commercial APIs or online services, where frequent queries can consume a substantial query budget.
Moreover, each query leaves an interaction trace on the server side, such as service-side logs or abnormal-access monitoring signals, thereby increasing query-induced exposure.
Figure~\ref{fig:motivation} shows the contrast between many and fewer queries in terms of query cost and query-induced exposure.
In particular, not all candidate samples are equally suitable for membership inference: some exhibit clear member/non-member contrast and provide strong membership signals, whereas many others yield only weak membership signals even after being queried~\cite{song2021systematic,carlini2022membership}.
Therefore, indiscriminately querying all candidate samples may waste a large fraction of the query budget on samples with limited benefit.
Together, these considerations raise a key question: which candidate samples are worth querying for black-box MIAs?

To address this question, we propose PSS-MIA, a pre-query sample selection framework for black-box MIAs that can be integrated with existing MIAs. PSS-MIA proceeds in two stages. The first stage ranks candidate samples without querying the target model and selects a subset expected to support stronger membership inference. The second stage applies a black-box MIA only to the selected samples. By limiting membership inference to the selected subset, PSS-MIA reduces query cost and query-induced exposure. In the ranking stage, we propose Loss-Gap Ranking (LGR).
LGR compares the losses of out-reference models (i.e., reference models trained without the candidate sample) and in-reference models (i.e., reference models trained with the candidate sample) to estimate membership signals, which serve as the basis for pre-query sample selection.

For well-generalized target models, black-box MIAs may exhibit weak attack performance over the full candidate set and a low full-set AUC may therefore underestimate their membership leakage risk. Nevertheless, the same attack may achieve stronger performance on a selected subset of candidate samples. To characterize the risk that is not fully reflected by the full-set AUC, we introduce a new metric called Target Attack-Performance Coverage (TAPC).

Given an attack goal, such as a specified AUC or ACC threshold, TAPC reports the size of the largest subset of candidate samples on which the evaluated MIA method achieves the attack goal. A positive TAPC indicates that at least one subset satisfies the prescribed attack goal, revealing membership leakage risk within part of the candidate set. A larger TAPC indicates that the goal is satisfied on a larger subset, reflecting more severe membership leakage. In some cases, AUC or ACC measured over the full candidate set may be low, but TAPC can expose substantial membership leakage risk on subsets.

\smallskip
\noindent\textit{In summary, our main contributions are as follows:}
\begin{itemize}[leftmargin=1.2em, topsep=2pt, itemsep=2pt, parsep=0pt, partopsep=0pt]
    \item We propose PSS-MIA, a pre-query sample selection framework that can be used with existing black-box MIAs. It ranks candidate samples, selects a subset expected to support stronger membership inference and only queries the target model with the selected subset, thereby reducing query cost and query-induced exposure.

    \item We propose LGR, which ranks candidate samples by estimating the strength of their membership signal using loss gaps computed from reference models. This ranking enables PSS-MIA to select samples on which the chosen black-box MIA achieves stronger member/non-member separability.
    
    \item We define a new metric called TAPC, which can evaluate the risk of models more reasonably. Instead of evaluating the privacy risk by an average metric over the full candidate set, TAPC focuses on the subsets that contain high-risk samples.

    \item Extensive experiments across multiple datasets, target architectures and representative black-box MIAs show that PSS-MIA with LGR consistently improves attack performance on selected subsets over baseline methods. Under low-FPR constraints, it also substantially reduces the target-query budget required to find fixed true positive samples. The results further identify a stable tail on which black-box MIAs show weak membership inference performance.
    
\end{itemize}
\section{Related Work}
\label{sec:related_work}

\subsection{Membership Inference Attacks}

Black-box MIAs aim to determine whether a candidate sample was a part of target model's training set, using only its accessible prediction outputs. Shokri et al.~\cite{shokri2017membership} pioneered the shadow-model paradigm, where auxiliary models mimic the target model and generate prediction outputs for training an attack classifier. Salem et al.~\cite{salem2019ml} demonstrated that such attacks remain effective even with fewer shadow models and less knowledge of the target model and its training data distribution. Truex et al.~\cite{truex2021demystifying} further studied membership inference in machine learning as a service (MLaaS) and analyzed how target model and data properties affect attack performance. Collectively, these studies established the shadow-model-based paradigm for inferring membership from black-box prediction outputs.

Subsequent research showed that black-box MIAs can infer membership directly from prediction statistics, without training an attack classifier. Yeom et al.~\cite{yeom2018privacy} proposed using the true-label loss as a membership score, while Sablayrolles et al.~\cite{sablayrolles2019white} analyzed loss-based membership inference from a Bayes-optimal perspective. Song et al.~\cite{song2019privacy} introduced modified prediction entropy and examined membership leakage in adversarially robust models. Song and Mittal~\cite{song2021systematic} later evaluated attacks based on confidence, entropy, modified entropy, and loss. Hui et al.~\cite{hui2021practical} proposed a blind MIA based on differential comparisons. These studies showed that simple statistics derived from prediction outputs can provide effective membership signals.

Another line of work considers more restrictive black-box settings where the target model returns only limited output information. Choquette-Choo et al.~\cite{choquette2021label} showed that membership can be inferred in a label-only setting using prediction stability under input perturbations. Li and Zhang~\cite{li2021membership} further proposed two decision-based attacks within this label-only setting: a transfer attack and a boundary attack. More recently, Li et al.~\cite{li2025enhanced} proposed DHAttack, a label-only MIA that reduces the per-sample query cost for boundary-distance estimation. These studies show that black-box membership inference remains possible even when the target model returns only predicted labels.

A parallel line of work investigates how additional black-box observations or query variants affect membership inference. Liu et al.~\cite{liu2024please} examined membership leakage when model explanations are exposed alongside prediction outputs. Wen et al.~\cite{wen2023canary} improved the performance of MIAs by querying adversarially optimized variants of the target sample. These studies focus on extracting richer membership signals from queried samples.

More recent work has used statistical testing and calibration to improve black-box MIAs. EnhancedMIA~\cite{ye2022enhanced} provides a hypothesis-testing framework that unifies several existing attacks. 
LiRA~\cite{carlini2022membership} trains shadow models with and without each candidate sample, estimates the corresponding score distributions, and then applies a likelihood-ratio test. Watson et al.~\cite{watson2022importance} adjust membership scores according to sample classification difficulty and show that this calibration can reduce false positives. Rezaei and Liu~\cite{rezaei2021difficulty} further show that many existing attacks struggle to achieve strong performance at low false positive rates. Quantile MIA~\cite{bertran2023scalable} uses quantile regression to estimate a sample-specific threshold on the confidence score. RMIA~\cite{zarifzadeh2024low} estimates membership using a likelihood-ratio test built from reference models and population samples, and remains effective with only a small number of reference models. RAPID~\cite{he2024difficulty} combines the original membership score with the calibrated score to reduce errors introduced by difficulty calibration.

Other attacks exploit training-process information. TrajectoryMIA~\cite{liu2022membership} uses the loss trajectory of each sample, and SeqMIA~\cite{li2024seqmia} evaluates multiple metrics on distilled models in chronological order and uses the resulting metric sequences for membership inference. These methods enrich the information used to compute membership scores. Different from all prior work, our work studies sample selection for black-box membership inference. We rank candidate samples and query only a selected subset, where existing black-box MIAs are expected to achieve stronger performance.

\subsection{Non-Uniform Membership Signals}

Membership-inference risk is not determined solely by a model's overall generalization behavior~\cite{aubinais2025fundamental}. Early analyses linked overfitting to membership leakage~\cite{yeom2018privacy}. Data augmentation and regularization can change MIA performance in ways that are not fully explained by the generalization gap~\cite{kaya2021does,tan2023blessing}. Consequently, models with similar overall generalization performance may still exhibit substantially different membership signals across individual samples. These studies mainly examine the relationship between model-level generalization behavior and membership leakage.

To explain this variation, prior work has linked this phenomenon to sample difficulty, learning dynamics, and memorization. Difficulty-calibrated attacks adjust the membership score according to the classification difficulty of the target sample. Studies of example forgetting show that individual training examples undergo forgetting events at different frequencies during learning~\cite{toneva2019empirical}. Work on long-tailed data distributions shows that memorization of rare and atypical examples can be necessary for achieving close-to-optimal generalization error~\cite{feldman2020does,feldman2020neural}, while subsequent work shows that memorization can be localized to a small subset of neurons distributed across layers~\cite{maini2023can}. These studies help explain why membership signals can vary considerably across samples, while their focus remains exclusively on the underlying causes and mechanisms.

Prior work has also examined membership leakage from other perspectives. Leino and Fredrikson~\cite{leino2020stolen} show that a model's idiosyncratic use of features can provide evidence of membership, and Kulynych et al.~\cite{kulynych2022disparate} studied differential vulnerability to MIAs across population subgroups. The privacy onion effect further shows that membership vulnerability can depend on the surrounding data distribution~\cite{carlini2022privacy}. These studies provide additional evidence that membership leakage is shaped by factors beyond global model accuracy.

Taken together, these studies show that membership signals can differ across individual samples due to generalization behavior, sample difficulty, memorization, feature use, subgroup disparity, and data composition. Our work builds on this observation to study a different problem: before target-model outputs are available, how an adversary can select candidate samples that are expected to enable stronger black-box membership inference. We address this problem by proposing PSS-MIA, which uses LGR to rank candidate samples before querying the target model. We further introduce TAPC to characterize membership leakage within candidate subsets.

\section{PSS-MIA}
\label{sec:method}

\begin{figure*}[!t]
    \centering
    \includegraphics[width=\textwidth]{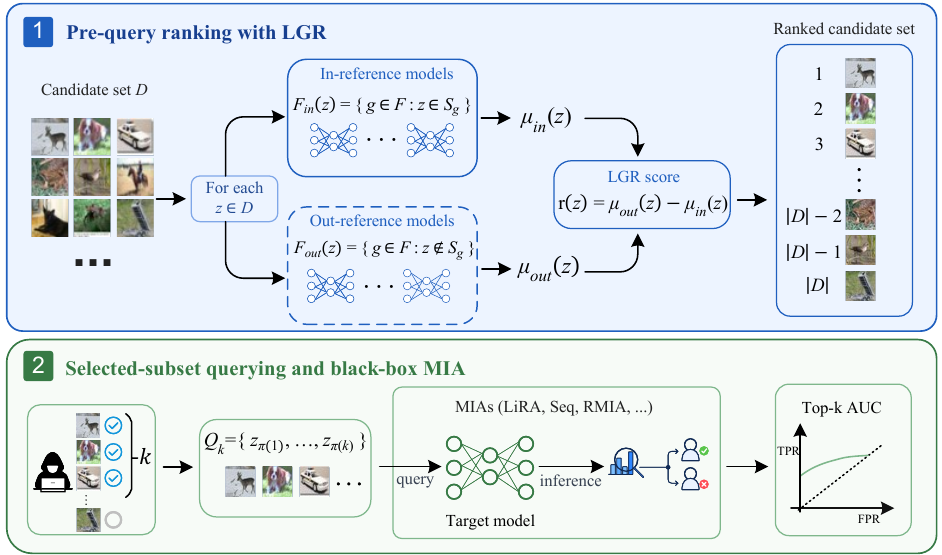}
    \caption{Overview of PSS-MIA. LGR ranks candidate samples using in-reference and out-reference losses before target model querying, after which an existing black-box MIA is applied to the selected samples.}
    \label{fig:framework}
\end{figure*}

\subsection{Problem Description and Threat Model}

We study a pre-query sample selection problem for black-box MIAs: before querying the target model, the adversary ranks candidate samples and then selects a subset expected to provide stronger membership signals.

Let $f$ denote the target model trained on a dataset $\mathcal{D}_{\mathrm{train}}$. Let
\[
D=\{z_i=(x_i,y_i)\}_{i=1}^{N}
\]
denote a candidate set for auditing that contains the member samples in $\mathcal{D}_{\mathrm{train}}$ together with additional non-member samples. For each $z_i \in D$, the black-box MIA aims to infer whether $z_i \in \mathcal{D}_{\mathrm{train}}$. Our focus is the preceding sample-selection stage. Specifically, we seek to construct a ranking $\pi$ over the candidate samples,
\[
z_{\pi(1)}, z_{\pi(2)}, \ldots, z_{\pi(N)},
\]
such that samples expected to exhibit stronger membership signals are ranked higher. Based on this ranking, the adversary can select the top samples for querying the target model, thereby reducing the query cost spent on candidates with weak membership signals and improving attack efficiency.

The threat model follows a label-aware black-box setting. The adversary knows the ground-truth label $y$ for each candidate sample $z=(x,y)\in D$, but does not know the exact composition of the target training set $\mathcal{D}_{\mathrm{train}}$ or which candidate samples are members. The adversary can access auxiliary data drawn from the target training distribution or from a closely matched distribution and can train reference models under protocols matched to the target setting. The pre-query selection stage is completed without querying the target model.

\subsection{Overview of the PSS-MIA}
\label{sec:pipeline}

PSS-MIA is a pre-query sample selection framework that can be integrated with existing black-box MIA methods. Figure~\ref{fig:framework} shows the overall workflow of PSS-MIA. Given a candidate set $D$, PSS-MIA first computes a score for each candidate sample without querying the target model. This score is not a membership prediction; rather, it estimates the expected strength of the sample's membership signal. Candidate samples are then ranked according to this score and a top subset is selected for the subsequent MIA. 

Let $r(z)$ denote the score for each candidate sample $z \in D$. For LGR, the score $r(z)$ is defined in Section~\ref{sec:lgr}. Sorting candidate samples in descending order of $r(z)$ yields a ranking $\pi$ such that
\[
r\bigl(z_{\pi(1)}\bigr)
\geq r\bigl(z_{\pi(2)}\bigr)
\geq \cdots
\geq r\bigl(z_{\pi(N)}\bigr).
\]
Given a selection size $k$, PSS-MIA selects the top subset
\[
Q_k=\{z_{\pi(1)},z_{\pi(2)},\dots,z_{\pi(k)}\}.
\]
The selection size $k$ determines the number of top samples selected for querying the target model and can be set according to the available query budget or a specified attack goal. These samples are expected to exhibit stronger membership signals.

After selecting $Q_k$, the adversary queries the target model only with the samples in $Q_k$ and obtains the corresponding target model outputs. These outputs are then used by an existing black-box MIA method $\mathcal{A}$ to infer membership. Let $s_{\mathcal{A}}(z)$ denote the membership score assigned by $\mathcal{A}$ to each $z \in Q_k$. PSS-MIA therefore changes only which candidate samples are queried, while leaving the chosen black-box MIA unchanged.

\subsection{Out-Minus-In Loss Shift as the LGR Ranking Signal}
\label{sec:loss_shift}

Before defining LGR, we provide an intuition for its ranking signal. For a candidate sample $z$, a model trained with $z$ is expected to incur a lower loss on $z$ than a model trained without $z$. Accordingly, a larger out-minus-in loss shift reflects a greater loss difference between the two cases, which may indicate a stronger membership signal. In the following analysis, we consider an ideal out-minus-in loss shift between models trained with and without $z$ and examine its relation to the Bayes error of in/out discrimination. This ideal quantity is used only to motivate the ranking signal; the practical LGR score is defined in Section~\ref{sec:lgr}.

For a candidate sample $z=(x,y)$ and a trained model $f$, we define the true-label loss as
\begin{equation}
\ell(f;z) = -\log p_f(y \mid x),
\label{eq:target_loss}
\end{equation}
where $p_f(y\mid x)$ denotes the predicted probability assigned by $f$ to the true label $y$.
Since this loss can be unbounded when $p_f(y\mid x)$ approaches zero, we introduce a clipped version:
\begin{equation}
\ell_B(f;z) = \min\{\ell(f;z), B\}, \qquad B>0.
\label{eq:clipped_loss}
\end{equation}
The clipping is used only to ensure boundedness in the analysis in this subsection. The practical LGR score in Section~\ref{sec:lgr} uses the unclipped loss in Eq.~\eqref{eq:target_loss}.

For this analysis, we consider the distributions of $\ell_B(f;z)$ under the two training conditions $z\in \mathcal{D}_{\mathrm{train}}$ and $z\notin \mathcal{D}_{\mathrm{train}}$. Let
\begin{equation}
\begin{aligned}
P_z^{\mathrm{in}}
&= \mathcal{L}\!\left(\ell_B(f;z)\mid z\in \mathcal{D}_{\mathrm{train}}\right),\\
P_z^{\mathrm{out}}
&= \mathcal{L}\!\left(\ell_B(f;z)\mid z\notin \mathcal{D}_{\mathrm{train}}\right).
\end{aligned}
\label{eq:conditional_laws}
\end{equation}
denote the corresponding loss distributions. The randomness comes from the training process in each case. We define the target-side out-minus-in mean loss shift as
\begin{equation}
\begin{aligned}
\Delta_B(z)
= {}& \mathbb{E}[\ell_B(f;z)\mid z\notin \mathcal{D}_{\mathrm{train}}] \\
&-\mathbb{E}[\ell_B(f;z)\mid z\in \mathcal{D}_{\mathrm{train}}].
\end{aligned}
\label{eq:target_shift}
\end{equation}

Under this definition, a positive $\Delta_B(z)$ indicates that the expected clipped loss of $z$ is higher under the out condition than under the in condition. 

Let $L_z^{\mathrm{in}}\sim P_z^{\mathrm{in}}$ and
$L_z^{\mathrm{out}}\sim P_z^{\mathrm{out}}$ denote the clipped-loss
variables under the two training conditions and define
\[
U_z^{\mathrm{in}}=\frac{L_z^{\mathrm{in}}}{B},
\qquad
U_z^{\mathrm{out}}=\frac{L_z^{\mathrm{out}}}{B}.
\]
We use $\operatorname{TV}(P,Q)$ to denote the total variation distance between two distributions $P$ and $Q$.
Then,
\begin{align}
\frac{\Delta_B(z)}{B}
&=
\mathbb{E}\!\left[U_z^{\mathrm{out}}\right]
-
\mathbb{E}\!\left[U_z^{\mathrm{in}}\right]
\\
&=
\int_0^1
\left(
\Pr\!\left(U_z^{\mathrm{out}}>t\right)
-
\Pr\!\left(U_z^{\mathrm{in}}>t\right)
\right)dt
\\
&\le
\int_0^1
\left|
\Pr\!\left(U_z^{\mathrm{out}}>t\right)
-
\Pr\!\left(U_z^{\mathrm{in}}>t\right)
\right|dt
\\
&\le
\operatorname{TV}\!\left(
P_z^{\mathrm{in}},P_z^{\mathrm{out}}
\right).
\end{align}

It follows that
\begin{equation}
\Delta_B(z)\le B\cdot \mathrm{TV}(P_z^{\mathrm{in}},P_z^{\mathrm{out}}).
\label{eq:delta_tv_main}
\end{equation}

Under equal priors, the Bayes error for discriminating between $P_z^{\mathrm{in}}$ and $P_z^{\mathrm{out}}$ based only on the clipped loss is
\begin{equation}
e_z^\star
=
\frac{1}{2}\Bigl(1-\mathrm{TV}(P_z^{\mathrm{in}},P_z^{\mathrm{out}})\Bigr),
\label{eq:bayes_error_main}
\end{equation}
which, together with Eq.~\eqref{eq:delta_tv_main}, implies
\begin{equation}
e_z^\star \le \frac{1}{2} - \frac{\Delta_B(z)}{2B}.
\label{eq:bayes_bound_main}
\end{equation}
This bound relates the signed out-minus-in mean loss shift to the Bayes error for distinguishing between the clipped-loss distributions under the in and out conditions. In particular, a larger positive $\Delta_B(z)$ gives a smaller upper bound on this Bayes error. 
A smaller Bayes error indicates greater separability between the clipped-loss distributions under the in and out conditions. This observation motivates using the out-minus-in loss shift as a ranking signal; it does not establish the optimality of the induced ranking.

The shift $\Delta_B(z)$ compares two target-side training cases for the same candidate sample, corresponding to whether $z$ is used for training. This shift is not available in practice and the pre-query ranking stage cannot rely on target model queries. To obtain a practical ranking signal, LGR uses matched reference models instead. It compares the losses of reference models trained with $z$ and those trained without $z$, using this reference-model loss gap to approximate the ideal shift $\Delta_B(z)$. The LGR score is defined in Section~\ref{sec:lgr}.

\subsection{Loss-Gap Ranking}
\label{sec:lgr}

Motivated by the analysis in Section 3.3, LGR uses reference models to construct a practical ranking signal. For each candidate sample, LGR compares its losses from in-reference and out-reference models and uses the resulting loss gap to rank candidate samples.

For a reference model $g$ and a candidate sample $z=(x,y)$, LGR uses the unclipped true-label loss
\[
\ell(g;z)=-\log p_g(y\mid x),
\]
where $p_g(y\mid x)$ denotes the probability assigned by $g$ to the true label $y$.

Let $\mathcal{F}$ denote the collection of reference models. For each candidate sample $z$, we define
\begin{equation}
\begin{aligned}
\mathcal{F}_{\mathrm{in}}(z)
&= \{g \in \mathcal{F} : z \in S_g\}, \\
\mathcal{F}_{\mathrm{out}}(z)
&= \{g \in \mathcal{F} : z \notin S_g\}.
\end{aligned}
\end{equation}
These sets contain the reference models trained with and without $z$, respectively, where $S_g$ is the training set of reference model $g$.

Let
\[
M_{\mathrm{in}}(z)=\lvert\mathcal{F}_{\mathrm{in}}(z)\rvert,
\qquad
M_{\mathrm{out}}(z)=\lvert\mathcal{F}_{\mathrm{out}}(z)\rvert.
\]
We assume that $M_{\mathrm{in}}(z)>0$ and $M_{\mathrm{out}}(z)>0$ for every candidate sample $z$.
The mean in-reference and out-reference losses are then computed as
\begin{equation}
\begin{aligned}
\mu_{\mathrm{in}}(z)
&=
\frac{1}{M_{\mathrm{in}}(z)}
\sum_{g\in\mathcal{F}_{\mathrm{in}}(z)}
\ell(g;z), \\
\mu_{\mathrm{out}}(z)
&=
\frac{1}{M_{\mathrm{out}}(z)}
\sum_{g\in\mathcal{F}_{\mathrm{out}}(z)}
\ell(g;z).
\end{aligned}
\end{equation}

The LGR score is then defined as the out-minus-in reference loss gap:
\begin{equation}
r(z)
=
\mu_{\mathrm{out}}(z)
-
\mu_{\mathrm{in}}(z).
\label{eq:lgr_score}
\end{equation}
A larger $r(z)$ indicates a stronger loss contrast between the out-reference and in-reference models.

LGR ranks candidate samples in descending order of $r(z)$ and selects the top $k$ samples for subsequent black-box membership inference.

LGR is a pre-query ranking method rather than a standalone MIA. Because $r(z)$ is computed entirely from reference-model losses, the ranking is obtained before the target model is queried on the candidate samples. After a top subset is selected, the target model is queried only on that subset and the returned outputs are used by the chosen black-box MIA.

\subsection{Target Attack-Performance Coverage}
\label{sec:tapc}

Conventional MIA evaluation often reports AUC, ACC, or TPR at a fixed FPR over the entire candidate set, but they do not reveal how membership signals vary across samples.
Therefore, weak full-set attack performance does not imply uniformly weak membership signals; a subset of samples may still provide sufficiently strong membership signals for effective membership inference.
To characterize the risk potentially obscured by full-set attack performance, we introduce the metric TAPC.
Given the ranked set of candidates $Q_N=\{z_{\pi(1)},z_{\pi(2)},\ldots,z_{\pi(N)}\}$ and an attack goal (e.g., $\mathrm{AUC}>0.90$), TAPC reports the largest subset size $k$ for which the chosen black-box MIA satisfies the attack goal on $Q_k$.

Let
\[
\mathcal{K}_{\mathrm{eval}}
=
\left\{
k_1,k_2,\ldots,k_m
\right\},
\qquad
k_1<k_2<\cdots<k_m,
\]
denote the ordered set of evaluated subset sizes.
Let $\mathcal{M}$ be an MIA evaluation metric for which larger values indicate stronger attack performance. Given a prescribed performance level $\tau$, we define
\begin{equation}
\mathrm{TAPC}_{\mathcal{M}}(\tau)
=
\max
\left\{
k\in\mathcal{K}_{\mathrm{eval}}
:
\mathcal{M}(Q_k)\geq\tau
\right\}.
\label{eq:tapc_general}
\end{equation}
If no evaluated subset satisfies the prescribed performance level, we set
\[
\operatorname{TAPC}_{\mathcal{M}}(\tau)=0.
\]
In our experiments, we use AUC and ACC as $\mathcal{M}$.

Under a fixed attack and evaluation protocol, $\operatorname{TAPC}_{\mathcal{M}}(\tau)>0$ indicates that at least one evaluated subset satisfies the prescribed performance level, revealing membership leakage risk that is concentrated in a subset of the candidate samples.
A larger TAPC indicates that the chosen black-box MIA satisfies the prescribed attack goal on a larger selected subset, revealing membership leakage over more candidate samples.
TAPC thus complements full-set attack performance by revealing membership-inference risk that may be concentrated among a subset of candidate samples.

\section{Experiments}
\label{sec:experiments}

We evaluate PSS-MIA with LGR across datasets, model architectures and MIAs. We measure attack performance on the selected subsets and target-query efficiency under low-FPR constraints. We also study the effect of subset size, the sensitivity of LGR to the number of reference model pairs and the tail produced by LGR. Finally, we use TAPC to measure the largest subset that satisfies a prescribed attack goal. Together, these experiments evaluate the effectiveness of PSS-MIA and LGR.

\begin{table*}[!t]
\centering
\caption{Attack performance (AUC, TPR@0.1\%FPR) under different ranking methods at 20\% selection ratio.}
\label{tab:fixed_k20_datasets}
\setlength{\tabcolsep}{3.5pt}
\renewcommand{\arraystretch}{1.08}
\resizebox{\textwidth}{!}{%
\begin{tabular}{cccccccccccc}
\toprule
\multirow{2}{*}{Dataset} & \multirow{2}{*}{Ranking method}
& \multicolumn{2}{c}{LiRA}
& \multicolumn{2}{c}{RMIA}
& \multicolumn{2}{c}{Seq}
& \multicolumn{2}{c}{Loss}
& \multicolumn{2}{c}{Attack\_R} \\
\cmidrule(lr){3-4}
\cmidrule(lr){5-6}
\cmidrule(lr){7-8}
\cmidrule(lr){9-10}
\cmidrule(lr){11-12}
& & AUC & TPR & AUC & TPR & AUC & TPR & AUC & TPR & AUC & TPR \\
\midrule

\multirow{4}{*}{CIFAR-10}
& Random & 74.20\% & 0.10 & 69.89\% & 0.03 & 76.30\% & 0.03 & 60.59\% & 0.00    & 69.62\% & 0.04 \\
& Loss   & 85.80\% & 0.18 & 79.69\% & 0.08 & 84.91\% & 0.06 & 71.10\% & 0.00    & 81.29\% & 0.11 \\
& LT-IQR    & 96.69\% & 0.29 & 88.43\% & 0.09 & 95.50\% & 0.11 & 92.03\% & 0.01 & 95.43\% & 0.24 \\
& \textbf{LGR} & \textbf{97.64\%} & \textbf{0.30} & \textbf{89.14\%} & \textbf{0.09} & \textbf{96.17\%} & \textbf{0.11} & \textbf{95.02\%} & \textbf{0.02} & \textbf{96.88\%} & \textbf{0.25} \\
\midrule

\multirow{4}{*}{CIFAR-100}
& Random & 94.78\% & 0.30 & 89.54\% & 0.24 & 96.38\% & 0.21 & 82.61\% & 0.00    & 91.16\% & 0.18 \\
& Loss   & 98.98\% & 0.54 & 96.62\% & 0.40 & 99.11\% & 0.22 & 94.38\% & 0.00   & 97.68\% & 0.30 \\
& LT-IQR    & 99.96\% & 0.89 & 99.66\% & 0.60 & 99.82\% & 0.60 & 99.70\% & 0.32 & 99.85\% & 0.55 \\
& \textbf{LGR} & \textbf{99.99\%} & \textbf{0.96} & \textbf{99.88\%} & \textbf{0.68} & \textbf{99.98\%} & \textbf{0.97} & \textbf{99.97\%} & \textbf{0.87} & \textbf{99.97\%} & \textbf{0.97} \\
\midrule

\multirow{4}{*}{CINIC-10}
& Random & 85.87\% & 0.15 & 80.10\% & 0.10 & 96.95\% & 0.09 & 71.27\% & 0.00    & 81.40\% & 0.12 \\
& Loss   & 93.22\% & 0.33 & 87.96\% & 0.22 & 99.36\% & 0.33 & 81.41\% & 0.00    & 90.17\% & 0.23 \\
& LT-IQR    & 99.36\% & 0.46 & 97.14\% & 0.26 & 99.34\% & 0.30 & 98.01\% & 0.01 & 98.91\% & 0.31 \\
& \textbf{LGR} & \textbf{99.65\%} & \textbf{0.52} & \textbf{98.15\%} & \textbf{0.26} & \textbf{99.48\%} & \textbf{0.61} & \textbf{99.24\%} & \textbf{0.13} & \textbf{99.50\%} & \textbf{0.36} \\
\bottomrule
\end{tabular}%
}
\end{table*}

\subsection{Experimental Setup}

Since there is no directly comparable framework for pre-query sample selection in black-box MIAs, we evaluate PSS-MIA by comparing different sample-ranking methods (random ranking, loss-based ranking, LT-IQR-based ranking and LGR) within the same framework proposed in this paper. We keep the overall PSS-MIA procedure fixed and vary only the ranking method. For each black-box MIA, all ranking methods use the same target model, candidate set and selection size.

Random ranking uses a random permutation of the candidate set. Loss-based ranking sorts candidate samples by their true-label loss from shadow models. LT-IQR-based ranking adapts the loss-trace interquartile-range method~\cite{pollock2025free} to our setting: we train shadow models using the candidate set, record the true-label loss trajectory of each candidate sample during shadow model training and rank samples by the interquartile range of the loss trajectory. We employ LiRA~\cite{carlini2022membership}, RMIA~\cite{zarifzadeh2024low}, Seq~\cite{li2024seqmia}, Loss attack~\cite{yeom2018privacy} and Attack\_R~\cite{ye2022enhanced} as black-box MIAs.

Given a size $k$, we first perform a global ranking over the candidate set without using target model outputs. We then select the highest-ranked $k/2$ members and $k/2$ non-members according to this ranking, ensuring that differences in AUC and TPR@FPR are not caused by class imbalance. Membership labels are used only to construct balanced subsets for controlled evaluation. They are not used by PSS-MIA during ranking, target model querying, or membership inference.

We report results from three perspectives: attack performance on selected subsets, TAPC and target-query efficiency. Top-$k$ AUC and TPR at low FPR measure attack performance on balanced evaluation subsets. TAPC measures the largest subset that satisfies a prescribed attack-performance level. 

Minimum Query Budget (MQB) is the minimum query budget required to attain a prescribed number of true positives under a fixed low-FPR constraint.
For MQB, we use the same balanced subset protocol described above and vary the selection size $k$. For a selected subset $Q_k$ and a low-FPR constraint $\alpha$, let
\[
\operatorname{TP}_{\alpha}(Q_k)
=
\max_{\theta:\operatorname{FPR}(Q_k,\theta)\le \alpha}
\operatorname{TP}(Q_k,\theta),
\]
where $\theta$ is the attack decision threshold and $\operatorname{TP}(Q_k,\theta)$ denotes the number of member samples in $Q_k$ correctly classified as members. Given a target number of true positives $m$, MQB is defined as
\[
\textnormal{MQB}_{\alpha}(m)
=
\min \{k : \operatorname{TP}_{\alpha}(Q_k) \ge m\}.
\]
If no evaluated subset reaches $m$ true positives, MQB is reported as not reached. A smaller MQB indicates that fewer target model queries are required to identify the same number of true-positive samples under the same low-FPR constraint and balanced evaluation protocol.

We evaluate PSS-MIA with LGR as its ranking method on CIFAR-10 and CIFAR-100~\cite{krizhevsky2009learning} and CINIC-10~\cite{darlow2018cinic} using three model architectures: ResNet-18~\cite{he2016deep}, VGG16~\cite{simonyan2015very} and WRN-28-2~\cite{zagoruyko2016wide}. For CIFAR-10 and CIFAR-100, we use the standard 50K images as the candidate set. For CINIC-10, we uniformly sample a fixed subset of 50K images and use it throughout the experiments. For each dataset, the candidate set is randomly divided into two halves. One half is used to train the target model and constitutes the member set, while the other half forms a disjoint non-member set. The resulting evaluation set therefore contains equal numbers of members and non-members. All target and reference models are trained using the same training configuration.

LGR uses 128 complementary reference-model pairs by default to compute the ranking scores. We also evaluate the sensitivity of LGR to the number of reference-model pairs.

\subsection{Attack Performance on Selected Subsets}
\label{sec:prefix_performance}

This subsection evaluates whether PSS-MIA with LGR achieves higher attack performance than using other ranking methods (random ranking, loss-based ranking and LT-IQR-based ranking). For each evaluated MIA, we keep the attack unchanged and vary only the ranking method and report AUC and TPR@0.1\%FPR on the resulting subsets.

Table~\ref{tab:fixed_k20_datasets} compares the attack performance of PSS-MIA under different ranking methods at a fixed selection ratio of 20\% across the three datasets, with ResNet-18 as the target architecture. Loss-based ranking, LT-IQR and LGR all outperform the random ranking, demonstrating the effectiveness of sample selection. Among these ranking methods, LGR achieves the best results across all reported settings. On CIFAR-10, for example, under LiRA, replacing random ranking with LGR increases AUC from 74.20\% to 97.64\% and TPR@0.1\%FPR from 0.10 to 0.30.
Notably, although LT-IQR achieves AUC values comparable to those of LGR in several settings, LGR yields substantially higher TPR@0.1\%FPR. For example, on CIFAR-100, LGR increases the TPR from 0.60 to 0.97 for Seq and from 0.32 to 0.87 for the Loss attack.

\begin{figure*}[!t]
    \centering
    \includegraphics[width=\textwidth]{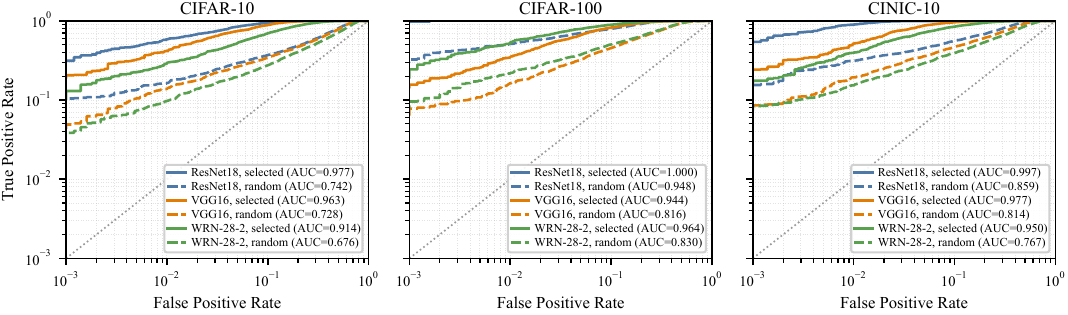}
    \caption{ROC comparison of LiRA across datasets and target architectures (LGR-selected vs. random).}
    \label{fig:roc_fixed_k20}
\end{figure*}

We further evaluate PSS-MIA across different target architectures. For each dataset, Figure~\ref{fig:roc_fixed_k20} reports the ROC curves of LiRA on ResNet-18, VGG16 and WRN-28-2, comparing PSS-MIA using LGR with PSS-MIA using random ranking. Across all the datasets and target architectures, PSS-MIA with LGR yields better ROC curves than PSS-MIA with random ranking.
\begin{table}[!t]
\centering
\caption{Attack performance (AUC/TPR@0.1\%FPR) at different selection sizes.}
\label{tab:topk_auc_prefix}
\setlength{\tabcolsep}{4pt}
\renewcommand{\arraystretch}{1.10}
\begin{tabular*}{\linewidth}{@{\extracolsep{\fill}}cccc@{}}
\toprule
size $k$ & LiRA & Loss & Seq \\
\midrule
5\%   & 99.77\% / 0.59 & 99.42\% / 0.20 & 99.39\% / 0.26 \\
10\%  & 99.42\% / 0.51 & 98.56\% / 0.10 & 98.71\% / 0.18 \\
20\%  & 97.64\% / 0.30 & 95.02\% / 0.02 & 96.17\% / 0.11 \\
50\%  & 87.62\% / 0.17 & 78.22\% / 0.00 & 87.62\% / 0.06 \\
100\% & 74.08\% / 0.09 & 60.28\% / 0.00 & 76.13\% / 0.03 \\
\bottomrule
\end{tabular*}
\end{table}

\subsection{Sensitivity to Selection Size and Reference-Model Pairs}
\label{sec:prefix_sensitivity}

We first examine how the attack performance of PSS-MIA changes as the selected subset size increases. Table~\ref{tab:topk_auc_prefix} reports top-$k$ AUC and TPR@0.1\%FPR for representative attacks on CIFAR-10 with ResNet-18. Across the reported attacks, both metrics decrease as $k$ increases. For LiRA, AUC decreases from 99.77\% at $k=5\%$ to 74.08\% at $k=100\%$, while TPR@0.1\%FPR decreases from 0.59 to 0.09. Seq and the loss-based attack show the same trend. These results demonstrate the effectiveness of LGR. When the selected subset is small, the queried samples retain stronger membership signals and support substantially higher attack performance. As the selection size increases, the subset includes more samples with weaker membership signals and the measured attack performance gradually decreases toward the full-set level.

\begin{figure}[!t]
    \centering
    \includegraphics[width=\columnwidth]{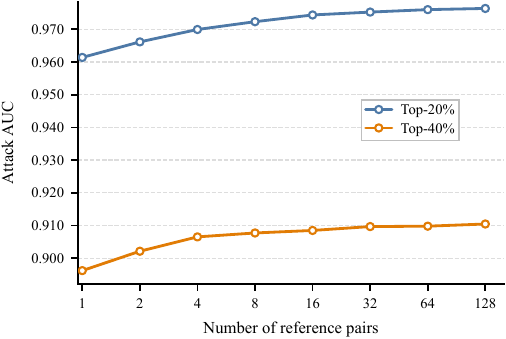}
    \caption{AUC on selected 20\% and 40\% candidate subsets with different numbers of reference-model pairs.}

    \label{fig:pair_sensitivity}
\end{figure}

We also examine the sensitivity of LGR to the number of reference model pairs used for score aggregation. We compute LGR scores with different numbers of complementary reference model pairs and evaluate the resulting rankings under the same selected-subset protocol. As shown in Figure~\ref{fig:pair_sensitivity}, using more reference-model pairs improves attack performance on the selected subsets, but the gain gradually diminishes. Increasing the number of complementary pairs from 1 to 128 raises the AUC on the selected 20\% subset from 0.961 to 0.976 and raises the AUC on the selected 40\% subset from 0.896 to 0.910. These results suggest that LGR already provides an effective ranking signal with a small number of reference-model pairs, while using more pairs brings further gains.

\subsection{Target-Query Cost and Query-Induced Exposure}
\label{sec:target_query_budget}

\begin{table*}[!t]
\centering
\caption{MQB of PSS-MIA at different low-FPR constraints and true positive targets.}
\label{tab:mqb_tp_all_datasets}
\setlength{\tabcolsep}{4.0pt}
\renewcommand{\arraystretch}{1.08}
\small
\resizebox{\textwidth}{!}{%
\begin{tabular}{llcccccccccccccccccc}
\toprule
\multirow{3}{*}[-0.4ex]{FPR} & \multirow{3}{*}[-0.4ex]{Rank.}
& \multicolumn{6}{c}{CIFAR-10}
& \multicolumn{6}{c}{CIFAR-100}
& \multicolumn{6}{c}{CINIC-10} \\
\cmidrule(lr){3-8} \cmidrule(lr){9-14} \cmidrule(lr){15-20}
& & \multicolumn{6}{c}{TP target $m$}
& \multicolumn{6}{c}{TP target $m$}
& \multicolumn{6}{c}{TP target $m$} \\
\cmidrule(lr){3-8} \cmidrule(lr){9-14} \cmidrule(lr){15-20}
& & 10 & 50 & 100 & 500 & 1000 & 2000
& 10 & 50 & 100 & 500 & 1000 & 2000
& 10 & 50 & 100 & 500 & 1000 & 2000 \\
\midrule
\multirow{3}{*}{$10^{-1}$}
& Rand. & 52 & 238 & 510 & 2638 & 5406 & 10914 & 24 & 138 & 266 & 1268 & 2524 & 5014 & 38 & 184 & 350 & 1808 & 3642 & 7234 \\
& LGR   & 20 & 100 & 200 & 1006 & 2016 & 4090  & 20 & 100 & 200 & 1000 & 2000 & 4000 & 20 & 100 & 200 & 1004 & 2000 & 4016 \\
& Red. ($\downarrow$)  & 61.5\% & 58.0\% & 60.8\% & 61.9\% & 62.7\% & 62.5\% & 16.7\% & 27.5\% & 24.8\% & 21.1\% & 20.8\% & 20.2\% & 47.4\% & 45.7\% & 42.9\% & 44.5\% & 45.1\% & 44.5\% \\
\midrule
\multirow{3}{*}{$10^{-2}$}
& Rand. & 72 & 572 & 1240 & 6014 & 11830 & 23820 & 42 & 202 & 374 & 1898 & 3858 & 7718 & 74 & 330 & 662 & 3344 & 6794 & 13254 \\
& LGR   & 20 & 100 & 200  & 1012 & 2050  & 4298  & 20 & 100 & 200 & 1000 & 2000 & 4002 & 20 & 100 & 200 & 1004 & 2000 & 4046 \\
& Red. ($\downarrow$)  & 72.2\% & 82.5\% & 83.9\% & 83.2\% & 82.7\% & 82.0\% & 52.4\% & 50.5\% & 46.5\% & 47.3\% & 48.2\% & 48.1\% & 73.0\% & 69.7\% & 69.8\% & 70.0\% & 70.6\% & 69.5\% \\
\midrule
\multirow{3}{*}{$10^{-3}$}
& Rand. & 118 & 1084 & 2098 & 10612 & 20976 & 41798 & 52 & 270 & 508 & 2834 & 5640 & 11496 & 102 & 624 & 1176 & 6004 & 11882 & 23866 \\
& LGR   & 20  & 100  & 200  & 1030  & 2140  & 5296  & 20 & 100 & 200 & 1000 & 2002 & 4016 & 20 & 100 & 200 & 1004 & 2040 & 4280 \\
& Red. ($\downarrow$)  & 83.1\% & 90.8\% & 90.5\% & 90.3\% & 89.8\% & 87.3\% & 61.5\% & 63.0\% & 60.6\% & 64.7\% & 64.5\% & 65.1\% & 80.4\% & 84.0\% & 83.0\% & 83.3\% & 82.8\% & 82.1\% \\
\bottomrule
\end{tabular}%
}

\end{table*}

Table~\ref{tab:mqb_tp_all_datasets} reports MQB of PSS-MIA across different low-FPR constraints and true positive targets. Unlike the fixed-subset evaluation in Section~4.2, MQB fixes a target number of true positive samples and measures the minimum subset size needed to find such a number of true positives. A smaller MQB indicates that PSS-MIA achieves this attack goal with a smaller target-query budget, providing a measure of target-query efficiency.

The results show that PSS-MIA with LGR substantially reduces MQB compared with the random-ranking baseline. For example, on CIFAR-10 under a $10^{-3}$ FPR constraint, random ranking requires 10{,}612, 20{,}976 and 41{,}798 target queries to find 500, 1{,}000 and 2{,}000 true-positive samples by LiRA, respectively, whereas PSS-MIA with LGR requires only 1{,}030, 2{,}140 and 5{,}296 queries. This yields reductions of 90.3\%, 89.8\% and 87.3\% in the target-query budget. These results indicate that PSS-MIA with LGR improves query efficiency: true positives under the low-FPR constraint appear earlier in the rank, so the same targets can be reached with fewer target queries.

This reduction is relevant beyond computational efficiency. In deployed black-box services, each target query is an observable interaction and may contribute to service-side logging, rate-limit pressure, or abnormal-access monitoring signals. Reducing unnecessary target queries can therefore lower target-query cost and may also reduce query-induced exposure.

\subsection{Stability and Weak Membership Signals in the Tail}
\label{sec:lower_ranked_tail}

The results in Section~\ref{sec:prefix_sensitivity} show that stronger attack performance is concentrated in the selected subset. We next examine the tail induced by LGR. Specifically, we test whether samples in this tail consistently show weak membership signals under the evaluated attacks and whether the tail remains stable across different LGR ranking trials.

We consider three analyses in this subsection: attack performance on tails, stability across different LGR ranking trials and retraining-based validation on the lower-ranked subset.

\begin{figure}[!t]
    \centering
    \includegraphics[width=\linewidth]{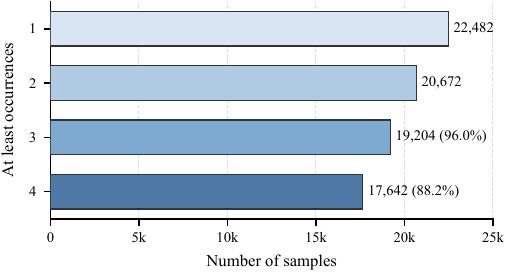}
    \caption{Occurrence frequency of samples in the 20{,}000-sample tails across different LGR ranking trials.}
    \label{fig:bottom_freq}
\end{figure}

\begin{table*}[!t]
\centering
\caption{Attack performance (AUC, TPR@0.1\%FPR) on top-ranked subsets and tails induced by LGR.}
\label{tab:top_bottom_subsets}
\small
\setlength{\tabcolsep}{4.8pt}
\renewcommand{\arraystretch}{1.10}
\begin{tabular*}{0.94\textwidth}{@{\extracolsep{\fill}}lcccccccccc@{}}
\toprule
\multirow{2}{*}{Ranked region}
& \multicolumn{2}{c}{LiRA}
& \multicolumn{2}{c}{RMIA}
& \multicolumn{2}{c}{Loss}
& \multicolumn{2}{c}{Seq}
& \multicolumn{2}{c}{Attack\_R} \\
\cmidrule(lr){2-3} \cmidrule(lr){4-5} \cmidrule(lr){6-7} \cmidrule(lr){8-9} \cmidrule(lr){10-11}
& AUC & TPR & AUC & TPR & AUC & TPR & AUC & TPR & AUC & TPR \\
\midrule
top-20\%    & 97.64\% & 0.30 & 89.14\% & 0.09 & 95.02\% & 0.02 & 96.17\% & 0.11 & 96.88\% & 0.25 \\
top-40\%    & 91.04\% & 0.19 & 82.59\% & 0.07 & 83.70\% & 0.00 & 90.48\% & 0.07 & 89.23\% & 0.23 \\
bottom-20\% & 51.15\% & 0.00 & 48.92\% & 0.00 & 50.14\% & 0.00 & 51.63\% & 0.00 & 49.49\% & 0.00 \\
bottom-40\% & 51.25\% & 0.00 & 50.61\% & 0.00 & 51.32\% & 0.00 & 54.30\% & 0.00 & 50.60\% & 0.00 \\
\bottomrule
\end{tabular*}
\end{table*}

Table~\ref{tab:top_bottom_subsets} compares the attack performance on top-ranked subsets and tails induced by LGR. The top-ranked subsets exhibit strong member/non-member separability, with LiRA reaching 97.64\% AUC on the top-20\% subset and 91.04\% AUC on the top-40\% subset. In contrast, the tails show much weaker membership signals: the AUC of LiRA drops to 51.15\% and 51.25\% on the bottom-20\% and bottom-40\% tails; the loss-based attack stays near chance and the TPR@0.1\%FPR is close to zero across attacks. These results indicate that the LGR-induced tails contain limited membership signals.

We next examine whether the tail remains stable across different LGR ranking trials, rather than being an artifact of a single run. To this end, we conduct four ranking trials, each of which trains 32 reference-model pairs independently. Figure~\ref{fig:bottom_freq} reports how many samples appear in at least 1, 2, 3, or 4 of the four tails: 19{,}204 samples appear in at least three tails, and 17{,}642 samples appear in all four tails, corresponding to 96.0\% and 88.2\% of a 20{,}000-sample tail, respectively. These results indicate that the tail induced by LGR is stable across different ranking trials.

\begin{figure}[!t]
    \centering
    \includegraphics[width=\linewidth]{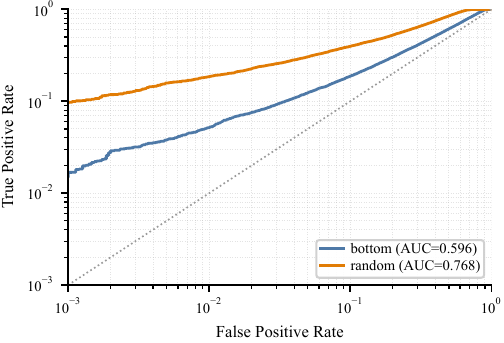}
    \caption{ROC curves of LiRA on target models retrained from the lower-ranked 30,000-sample subset and from a random 30,000-sample subset.}
    \label{fig:retrain_compare}
\end{figure}

To further characterize this phenomenon, we examine whether a lower-ranked subset also yields weaker membership signals when the target model is retrained. We take the lower-ranked 30{,}000-sample subset directly induced by LGR and randomly split it into two halves (using one half for training and the other half as a non-member set). As shown in Figure~\ref{fig:retrain_compare}, LiRA achieves an AUC of only 0.596 on the target model retrained from this lower-ranked subset, whereas it achieves a much higher AUC of 0.768 when the same retraining protocol is applied to a random 30{,}000-sample subset.

These results show that membership signals differ significantly between top-ranked subsets and the tails induced by LGR. This rank-wise variation can be obscured by full-set attack metrics, making it important to propose a new evaluation metric. We next use TAPC to measure the largest subset that satisfies a prescribed attack goal.

\subsection{Attack-Performance Coverage Beyond Full-Set Evaluation}

For well-generalized target models, stronger generalization can substantially weaken full-set membership inference performance~\cite{dionysiou2023sok}. However, weak full-set attack performance does not necessarily imply that membership signals are uniformly weak across the candidate set. To make this distinction explicit, we evaluate a well-generalized target model and use TAPC to characterize membership leakage risk that may be obscured by weak full-set attack performance.

We construct this target model by strengthening data augmentation and regularization, while keeping the remaining training and evaluation protocol unchanged. Under this setting, the target model has a generalization gap of about 3\%. With the MIA method kept unchanged, we rank candidate samples with LGR and compare LiRA performance on the full candidate set and on the subset selected by LGR.
\begin{figure}[!t]
    \centering
    \includegraphics[width=\linewidth]{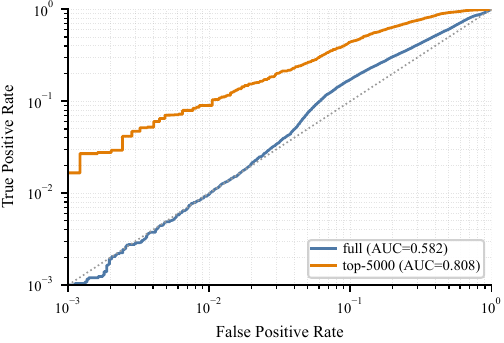}
    \caption{ROC curves of LiRA on the full candidate set and on the LGR-selected 5{,}000-sample subset for the well-generalized target model.}
    \label{fig:better_generalized_target}
\end{figure}

As shown in Figure~\ref{fig:better_generalized_target}, LiRA achieves an AUC of 0.582 on the full candidate set, but the AUC increases to 0.808 on the selected subset. This gap shows that weak full-set attack performance can obscure stronger membership signals within selected candidate subsets.

\begin{table}[!t]
\centering
\caption{TAPC comparison under different attack goals.}
\label{tab:tapc_generalized_vs_standard}
\renewcommand{\arraystretch}{1.10}
\begin{tabular*}{\linewidth}{@{\extracolsep{\fill}}ccc@{}}
\toprule
Attack goal & Well-generalized model & Standard model \\
\midrule
AUC $= 0.8$ & 5666  & 28882 \\
AUC $= 0.7$  & 17180 & 50000 \\
ACC $= 0.8$ & 40    & 17354 \\
ACC $= 0.7$  & 9344 & 37324 \\
\bottomrule
\end{tabular*}
\end{table}
Table~\ref{tab:tapc_generalized_vs_standard} further quantifies this difference using TAPC. Compared with the standard target model, the well-generalized target model has smaller TAPC values under the same AUC or ACC threshold, indicating that stronger generalization can reduce the membership leakage risk. Nevertheless, the TAPC values remain positive under several AUC and ACC thresholds. This shows that weak full-set attack performance alone does not rule out localized membership leakage within the candidate set.

\subsection{Cross-Architecture Generalization}

\begin{table*}[!t]
\centering
\caption{Cross-architecture evaluation of LGR (AUC/TPR@0.1\%FPR).}
\label{tab:cross_arch}
\renewcommand{\arraystretch}{1.20}
\setlength{\tabcolsep}{4pt}
\begin{tabular*}{0.86\textwidth}{@{\extracolsep{\fill}}ccccc@{}}
\hline
\multirow{2}{*}{Dataset} & \multirow{2}{*}{Reference architecture} & \multicolumn{3}{c}{Target model architecture} \\
\cline{3-5}
& & ResNet-18 & VGG16 & WRN-28-2 \\
\hline
\noalign{\vskip 1pt}
\multirow{3}{*}{CIFAR-10}
& ResNet-18 & 97.64\% / 0.30 & 94.86\% / 0.19 & 91.26\% / 0.13 \\
& VGG16    & 96.79\% / 0.29 & 96.25\% / 0.19 & 90.51\% / 0.13 \\
& WRN-28-2 & 97.07\% / 0.30 & 94.86\% / 0.18 & 91.44\% / 0.13 \\
\noalign{\vskip 2pt}
\hline
\noalign{\vskip 2pt}
\multirow{3}{*}{CIFAR-100}
& ResNet-18 & 99.99\% / 0.96 & 90.45\% / 0.11 & 95.86\% / 0.25 \\
& VGG16    & 99.79\% / 0.76  & 94.40\% / 0.16 & 94.58\% / 0.20 \\
& WRN-28-2 & 99.98\% / 0.95  & 91.62\% / 0.12 & 96.35\% / 0.24 \\
\noalign{\vskip 2pt}
\hline
\noalign{\vskip 2pt}
\multirow{3}{*}{CINIC-10}
& ResNet-18 & 99.65\% / 0.52 & 96.57\% / 0.23 & 94.90\% / 0.17 \\
& VGG16    & 99.31\% / 0.53 & 97.68\% / 0.24 & 94.31\% / 0.17 \\
& WRN-28-2 & 99.48\% / 0.54 & 96.54\% / 0.23 & 95.02\% / 0.17 \\
\noalign{\vskip 2pt}
\hline
\end{tabular*}
\end{table*}

We further evaluate whether the LGR ranking signal transfers across model architectures. Although the default setting uses reference models matched to the target setting, the adversary may not know the exact target architecture in practice. We train reference models using ResNet-18, VGG16, and WRN-28-2 to compute LGR scores. We then evaluate LiRA on the selected 20\% subsets for target models using each of the three architectures, covering both matched and mismatched architecture settings. The candidate set and selection ratio are kept fixed throughout. Table~\ref{tab:cross_arch} reports AUC/TPR@0.1\%FPR of PSS-MIA under different reference and target model architectures.

Table~\ref{tab:cross_arch} shows that LGR generally remains effective under different reference and target model architectures. The matched-architecture setting often achieves the best results, but mismatched reference architectures still produce strong results on the selected subsets. These results indicate that the out-minus-in reference loss gap provides a transferable ranking signal across model architectures.

\subsection{Ablation Study}

\begin{table}[!t]
\centering
\caption{Reference-loss signal ablation (AUC, TPR@0.1\%FPR).}
\label{tab:score_ablation}
\small
\setlength{\tabcolsep}{3.2pt}
\renewcommand{\arraystretch}{1.05}
\begin{tabular}{ccccccc}
\toprule
\multirow{2}{*}{Ranking score}
& \multicolumn{2}{c}{LiRA}
& \multicolumn{2}{c}{Seq}
& \multicolumn{2}{c}{Loss} \\
\cmidrule(lr){2-3} \cmidrule(lr){4-5} \cmidrule(lr){6-7}
& AUC & TPR & AUC & TPR & AUC & TPR \\
\midrule
In-reference 
& 93.69\% & 0.25 
& 95.52\% & 0.19 
& 84.96\% & 0.00 \\

Out-reference 
& 98.60\% & 0.41 
& 98.82\% & 0.22 
& 95.40\% & 0.00 \\

Out-minus-in
& 99.60\% & 0.51 
& 98.95\% & 0.38 
& 98.67\% & 0.03 \\
\bottomrule
\end{tabular}
\end{table}

We further examine whether LGR benefits from contrasting in-reference and out-reference losses, rather than relying on one-sided reference-loss signals alone. To this end, we conduct an ablation study with three scoring variants: (1) in-reference loss; (2) out-reference loss; (3) the out-minus-in loss gap used by LGR. The in-reference-loss score ranks candidate samples by their mean loss from in-reference models, while the out-reference-loss score ranks them by their mean loss from out-reference models. In contrast, LGR ranks candidate samples by the out-minus-in loss gap, which contrasts the mean out-reference loss with the mean in-reference loss for the same sample. To compute the in-reference-loss score, we use reference models that include the candidate sample in training; to compute the out-reference-loss score, we use reference models trained on a disjoint 25K CINIC-10 auxiliary subset, so the evaluated candidate samples are excluded from those models. For a fair comparison, we keep the target model, candidate set, MIA methods, and subset-selection protocol fixed across all scoring variants.

Table~\ref{tab:score_ablation} reports the ablation results. The two one-sided reference-loss scores still provide useful ranking signals. Nevertheless, ranking by the out-minus-in loss gap consistently achieves the best performance across the evaluated attacks. For LiRA, AUC/TPR@0.1\%FPR increases from 93.69\%/0.25 with in-reference loss and 98.60\%/0.41 with out-reference loss to 99.60\%/0.51 with the out-minus-in loss gap. Seq and the Loss attack exhibit the same phenomenon. These results suggest that contrasting in-reference and out-reference losses provides a more effective ranking signal than using either reference-loss signal alone.

\section{Limitations}

Our empirical evaluation focuses on image classification benchmarks, e.g., CIFAR-10, CIFAR-100, and CINIC-10. Although these datasets are commonly used in membership inference evaluation, whether PSS-MIA and LGR remain effective for other data modalities, tasks, and model types, such as language models and generative models, remains an open question.

LGR requires multiple reference models to estimate the out-minus-in loss gap. This requirement may limit its applicability in settings with limited computational resources or limited auxiliary data. At the same time, this requirement is comparable to the assumptions of MIAs based on reference or shadow models, such as LiRA and RMIA, which also require auxiliary models and data to estimate membership-related signals.

For controlled evaluation, we use balanced subsets containing equal numbers of members and non-members. This design makes different ranking methods directly comparable and avoids confounding changes in AUC or TPR@FPR caused by class imbalance. In realistic attack settings, however, the membership prior may be unknown or imbalanced, which can affect measured attack performance. Membership labels are used only to construct evaluation subsets after the ranking is fixed, and they are not used during ranking, target-model querying, or membership inference.
\section{Conclusion}

In this paper, we studied pre-query sample selection for black-box MIAs and proposed PSS-MIA. We further introduced LGR to rank candidate samples using reference-model loss gaps. Across datasets, target architectures, and black-box MIAs, PSS-MIA with LGR selects subsets on which existing attacks achieve stronger member/non-member separability. Since only the selected subset is queried, this selection process reduces the number of target-model queries required to reach fixed true-positive targets under low-FPR constraints.

Moreover, we proposed a new metric called TAPC to characterize membership leakage that may not be reflected by full-set evaluation, providing a complementary view of black-box MIA risk. Experimental results on well-generalized target models demonstrate the effectiveness of TAPC. These results highlight pre-query sample selection and TAPC as important perspectives for evaluating and understanding black-box membership inference risk.

\bibliographystyle{IEEEtran}
\bibliography{refs}

\end{document}